\documentclass[
reprint,
 amsmath,amssymb,
 aps,
]{revtex4-2}
\usepackage{graphicx}
\usepackage{dcolumn}
\usepackage{bm}
\usepackage[caption=false]{subfig}
\usepackage{url}
\usepackage{float}
\usepackage{subfloat}
\usepackage[export]{adjustbox}
\usepackage{multirow}
\usepackage{color}
\usepackage{fancyhdr}
\usepackage{longtable}
\usepackage{parskip}
\usepackage[T1]{fontenc}
\usepackage[utf8]{inputenc}
\usepackage{multibib}
\usepackage{natbib}
\usepackage{breqn}
\usepackage{comment}
\usepackage{soul}
\begin{document}
\title{Relative cluster entropy for power-law correlated sequences}
\author{Anna Carbone}
\affiliation{%
 Politecnico di Torino Italy
}%
\author{Linda Ponta}%
\affiliation{LIUC Castellanza Italy}
\date{\today}%
\begin{abstract}
We propose an information-theoretical measure, the \textit{relative cluster entropy}   $\mathcal{D_{C}}[P \| Q] $,  to discriminate among cluster partitions characterised by probability distribution functions $P$ and $Q$. The measure  is illustrated with the clusters generated by pairs of fractional Brownian motions with Hurst exponents $H_1$ and $H_2$   respectively.  For subdiffusive, normal and superdiffusive sequences, the relative entropy sensibly depends  on the difference between $H_1$ and $H_2$. By using  the \textit{minimum relative  entropy} principle,  cluster sequences characterized by different correlation degrees are distinguished  and the optimal Hurst exponent is selected.  As a case study, real-world  cluster partitions of  market price series  are compared to those obtained from  fully uncorrelated   sequences (simple Browniam motions) assumed  as a model. The \textit{minimum relative cluster entropy} yields optimal Hurst exponents $H_1=0.55$,  $H_1=0.57$,   and $H_1=0.63$  respectively for the prices of  DJIA, S\&P500, NASDAQ: a clear indication of non-markovianity. Finally, we derive the analytical expression of the relative cluster entropy and  the  outcomes are discussed for arbitrary pairs of power-laws probability distribution functions of continuous random variables.
\end{abstract}
\maketitle
\section{Introduction}
\label{sec:Introduction}
Flow of information in complex systems with interacting components can be quantified via entropy measures \cite{cafaro2016thermodynamic,parrondo2015thermodynamics,kawai2007dissipation,horowitz2014thermodynamics,still2012thermodynamics,ortega2013thermodynamics,san2005information}.
In this context, discriminating between empirical data and  models  in terms of information content is interesting from several viewpoints. Consider an experiment with the outcomes obeying the probability distribution $P$ whereas the distribution $Q$ is a model for the same experiment. Quantifying the error of the wrong assumption of the model compared to the empirical information content is relevant  to a broad class of  phenomena \cite{chen2021wiener,vedral2002role}. Such information-theoretical concepts bring also together the thermodynamic implications intrisically related to the evolution of the system under investigation. The dynamic of the information  transferred along  subsequent transformative states of a  complex system can be described  in terms of divergence of the probability distributions $P$ at time $t$ and $P'$ at a subsequent time $t'$.
Hence, information-theoretical tools finds applications   in fields as diverse as climate, turbulence, neurology, biology and economics \cite{kleeman2002measuring,granero2018kullback,backus2014sources,tozzi2021information} and are increasingly adopted  in unsupervised learning of unlabelled data where similarity/dissimilarity measures are concerned with dynamic rather than static features of the clustered data \cite{ullmann2021validation,meilua2007comparing,liao2005clustering}.
\par
A recently proposed information measure,  with the ability to quantify heterogeneity  and dynamics of long-range correlated processes in a broad range of  application areas, is the \textit{cluster entropy}  $\mathcal{S_{C}}[P]$ \cite{carbone2004analysis,carbone2007scaling,carbone2013information,ponta2021information}. The measure has been defined as a Shannon functional
 with  $P$  the power-law probability distribution of the clusters  formed in a long-range correlated data sets.  If $P$ is a distribution concentrated on a single cluster value,  $\mathcal{S_C} [P] = 0 $  corresponds to the minimum uncertainty  on the outcome of the cluster size,  the random variable of interest. If $P$ is a fully developed power-law distribution, $\mathcal{S_C}[P]=\ln N$ corresponds  to the maximum uncertainty obtained as the power-law  distribution spreads over a broad range of cluster values. Thus, according to the Shannon interpretation,  $\mathcal{S_C} [P]$ can be understood as a measure of uncertainty of all the possible cluster outcomes.  By extending the definition to continuous variables,  the \textit{differential cluster entropy} ${S_{C}}[P]$ added clues to the approach  by clarifying the interplay of the different terms entering the cluster entropy and thus  the  origin of the excess randomness.
\par
In this work, we go beyond the simple \textit{measure of uncertainty} of the random variable outcomes  provided by $\mathcal{S_C} [P]$.
An inference method for hypothesis testing of a general class of  models underlying a relevant stochastic process is developed.  We put forward the   \textit{relative cluster entropy}  or \textit{cluster divergence} $\mathcal{D_{C}}[P \| Q] $ with the first argument $P$  the empirical distribution and the second argument $Q$ a model  within a broad class of  probability distributions.  $\mathcal{D_{C}}[P \| Q] $  is  a metric on the space of probability distributions, interpreted as a divergence rather than as a  distance since it does not obey symmetry and triangle inequality.
The asymmetry of the relative entropy reflects the asymmetry between data and models, hence it can be used for inference purposes on the model underlying a given distribution.  If $\mathcal{D_{C}}[P \| Q]  >> 0 $,  the hypothesis likelihood is very low and, unless the quality of the empirical data  should be questioned, the model distribution $Q$ must be rejected. The higher $\mathcal{D_{C}}[P \| Q] $, the lower the likelihood of the hypothesis. If the hypothesis on the model were true, $P$ should fluctuate around its expected value $Q$, with  fluctuations of limited amplitude and occurrence probability greater than the significance level, resulting in the acceptance of the  model $Q$.
\par
The \textit{cluster entropy} $\mathcal{S_{C}}[P]$  and the \textit{relative cluster entropy} $\mathcal{D_{C}}[P \| Q] $ can be interpreted as information measures over partitions generated by  a coarse-grained mapping of the two-dimensional phase-space spanned by a particle, e.g. a simple Brownian path described by the random variable $\{x_t \}$. According to Gibbs’ original idea at the core of the information entropy concept, a coarse grained description is defined by smoothing out  fine details and increasing the observer’s ignorance about the exact microstate of the system.   As the structure description becomes blurrier, randomness and entropy increase. A coarse-grained  description is obtained by performing a local average over the phase-space cells with increasing size.
In the  information clustering approach adopted here, the coarse grained description of the particle path  $\{x_t \}$ is obtained by a local average $\{\widetilde{x}_{t,n}\}$ over the phase-space cells with the parameter $n$ defining the cell sizes.  The linear regression
$ {x_{t}}= \widetilde{x}_{t,n}+\epsilon_{t,n} $ yields the errors $\epsilon_{t,n}={x_{t}}- \widetilde{x}_{t,n}$ which ultimately generate a finite partition $\left\{\mathcal{C}\right\}=\left\{\mathcal{C}_{n, 1}, \, \mathcal{C}_{n, 2}, \, \ldots , \, \mathcal{C}_{n, j}\right\}$ for each $n$. The  partition process generate regions, named as \textit{clusters}, bounded between the values of $t$ when $\epsilon_{t,n}=0$, which correspond to complete information with minimum entropy.  The probability distribution functions of the random variables defined by $\epsilon_{t,n} $ univoquely quantify the loss of structure and information of the coarse grained representation.
As already noted, the \textit{cluster entropy} $\mathcal{S_{C}}[P]$ is bounded,  involves integrating over cell components, ranges from the minimum  to the maximum value  as the  description ranges from the finest-grained (corresponding to the smallest clusters) to the coarsest-grained partition (corresponding to the largest clusters).
The \textit{relative cluster entropy} $\mathcal{D_{C}}[P \| Q] $ is also bounded, involves integration over cells, ranges   between a maximum value, depending on the  two distributions,  and the minimum value $0$ for $P=Q$.
\par
The ability of the cluster divergence $\mathcal{D_{C}}[P \| Q] $ to select an optimal distribution could be relevant in several contexts.  In particular, complex phenomena obeying power-law distributions are still raising concerns regarding accuracy and veracity of the estimation  of the power law exponent \cite{clauset2009power}.
To illustrate how  the \textit{relative cluster entropy} operates, synthetic and real-world  data featuring power-law distribution behavior  are considered. First, the approach is implemented on pairs of synthetic fractional Brownian motions ({\em fBms}) with  given Hurst exponent. A systematic dependence of $\mathcal{D_{C}}[P \| Q] $  on the Hurst exponents of the pair is found.  The \textit{minimum relative entropy}   principle is then implemented  as a selection criterion to extract  the optimal correlation exponent of the sequence. Second, as a real-world case, we study  the divergence $\mathcal{D_{C}}[P \| Q] $ of financial price series. The probability distribution $P$ is obtained by ranking the clusters generated in each price time series and compared to the distribution $Q$ drawn from synthetic {\em fBms} data adopted as model. The \textit{minimum relative entropy} principle yields  the best estimate of the correlation exponents of the  financial series and quantifies the deviation of the price series from the assumed model.
\par
The manuscript is organized as follows. In Section \ref{sec:Relative} the main computational steps of the \textit{relative cluster entropy} method are described for  discrete variables. The approach is illustrated for synthetic  (fractional Brownian motions) and real-world  (market price series) data.
In Section \ref{sec:Discussion} the \textit{relative cluster entropy} is extended to continuous random variables, conclusions and suggestions for further developments are drawn.
\begin{figure*}[htb!]
\includegraphics[scale=0.3]{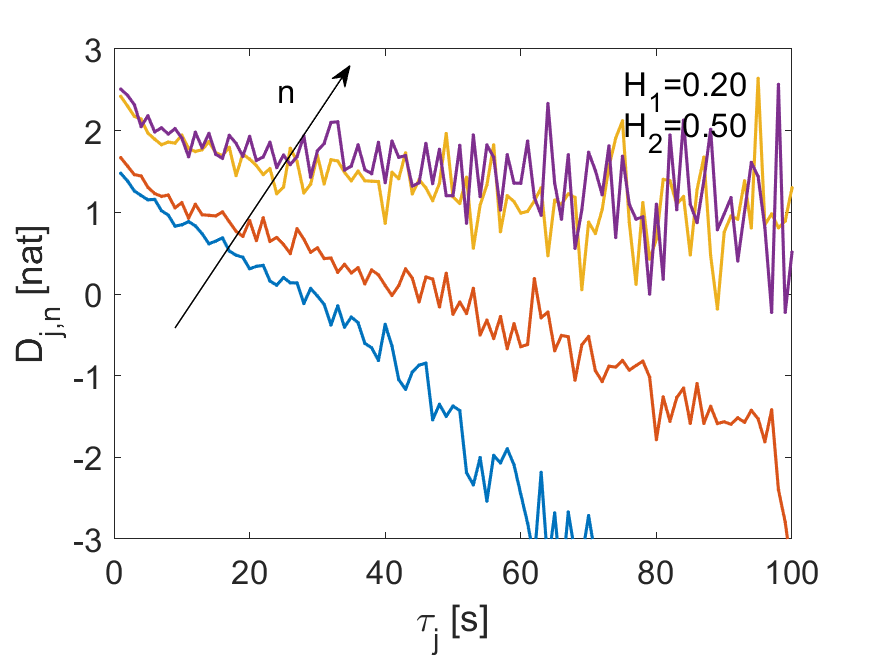}
\includegraphics[scale=0.3]{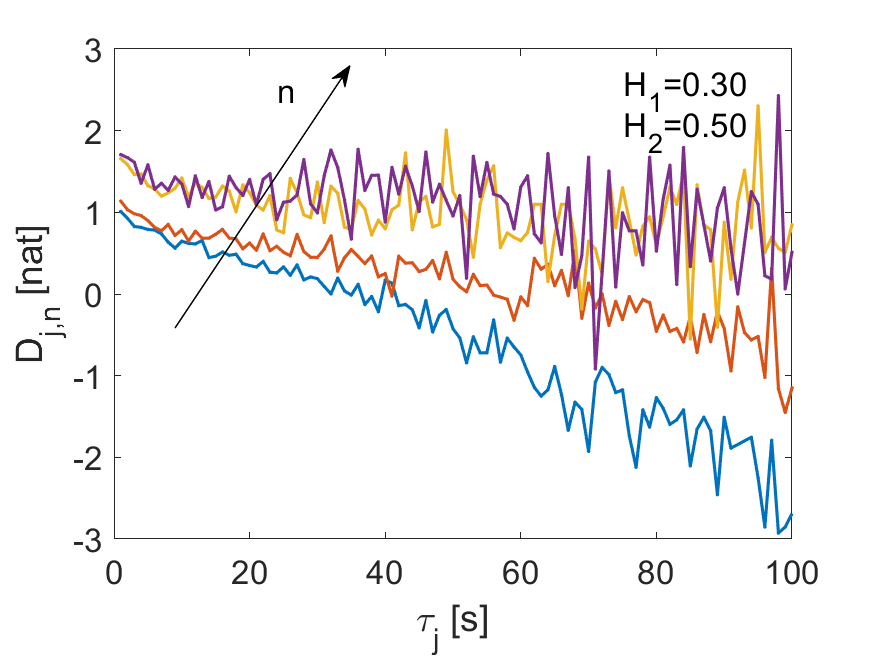}
\includegraphics[scale=0.3]{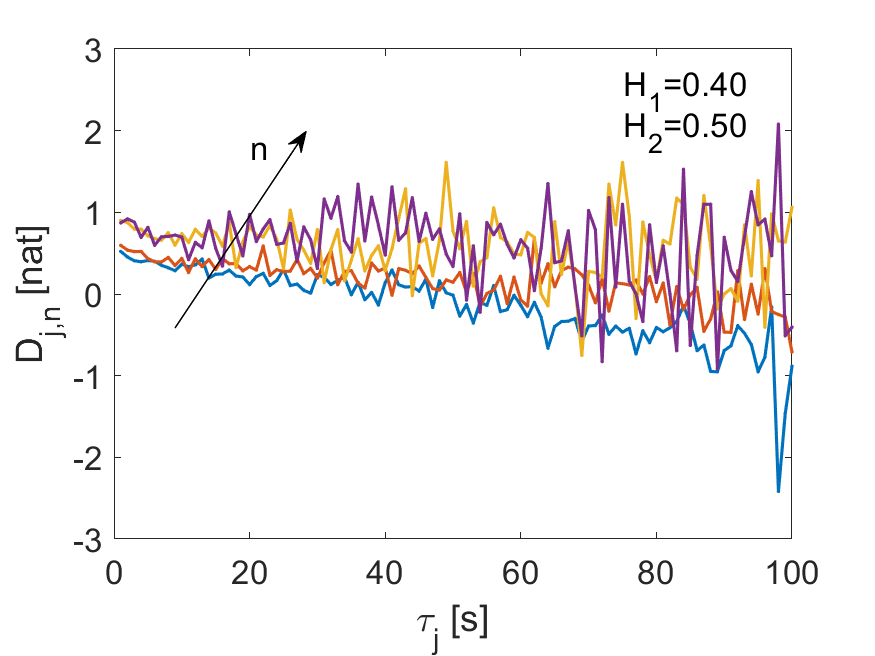}\\
\includegraphics[scale=0.3]{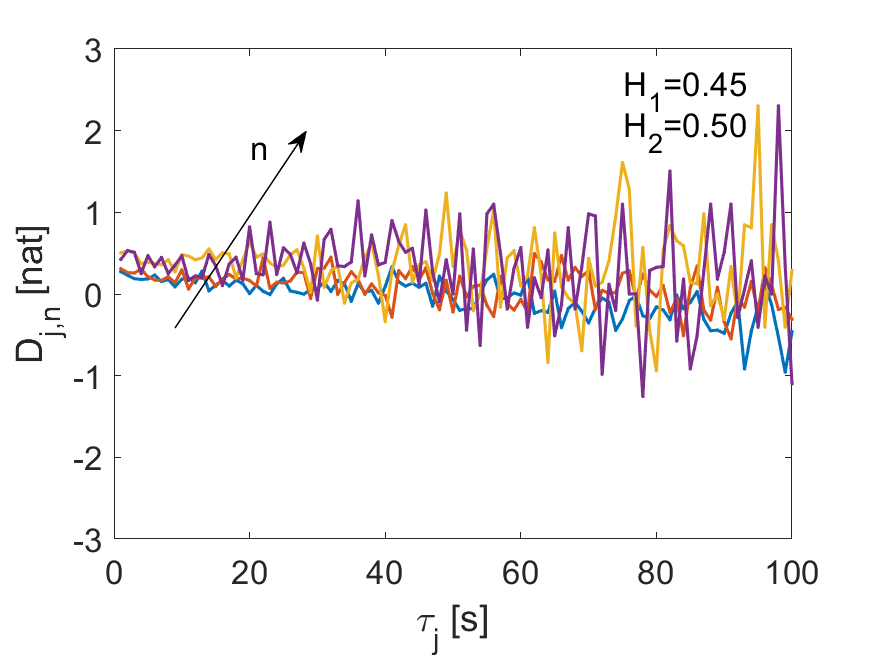}
\includegraphics[scale=0.3]{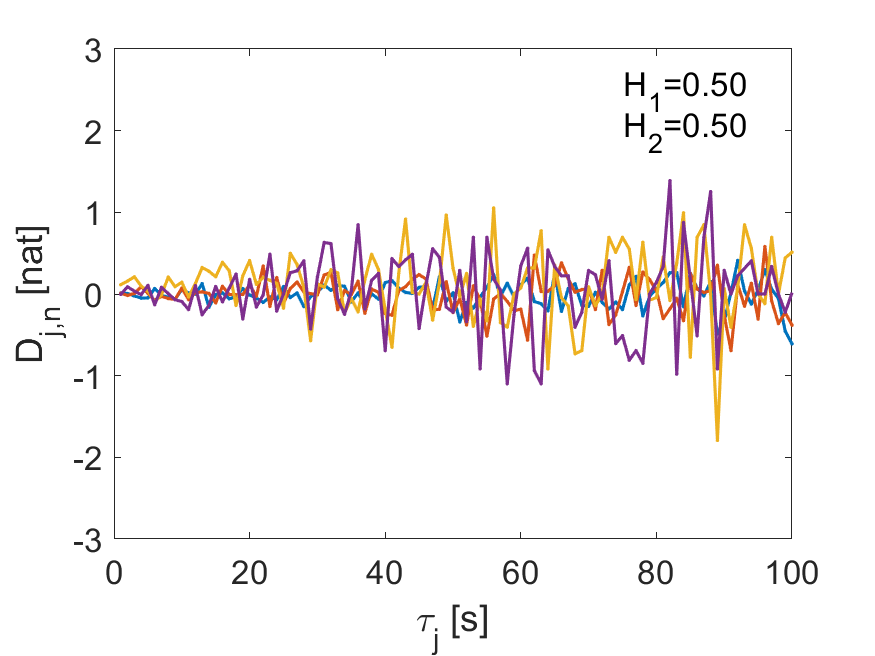}
\includegraphics[scale=0.3]{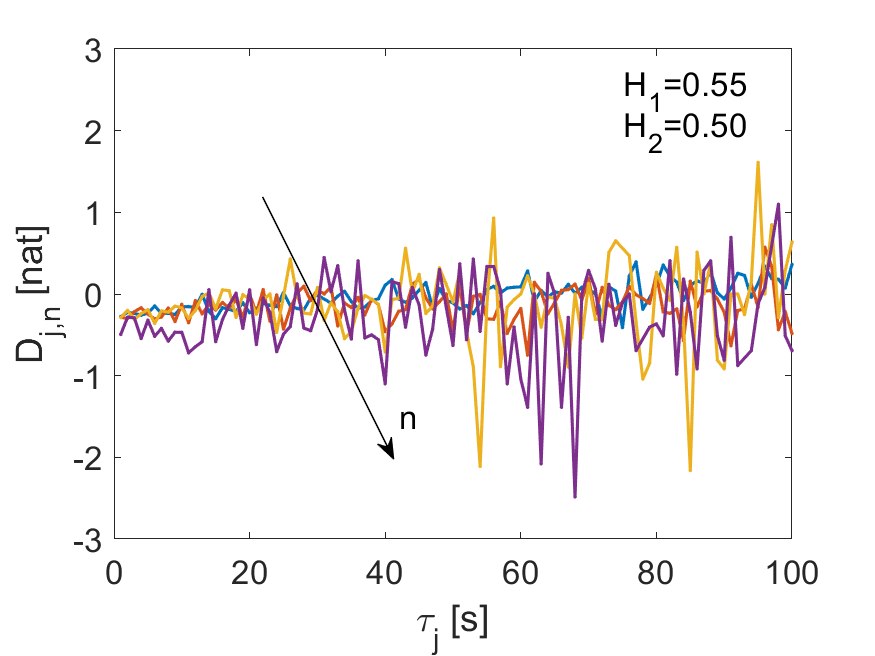}\\
\includegraphics[scale=0.3]{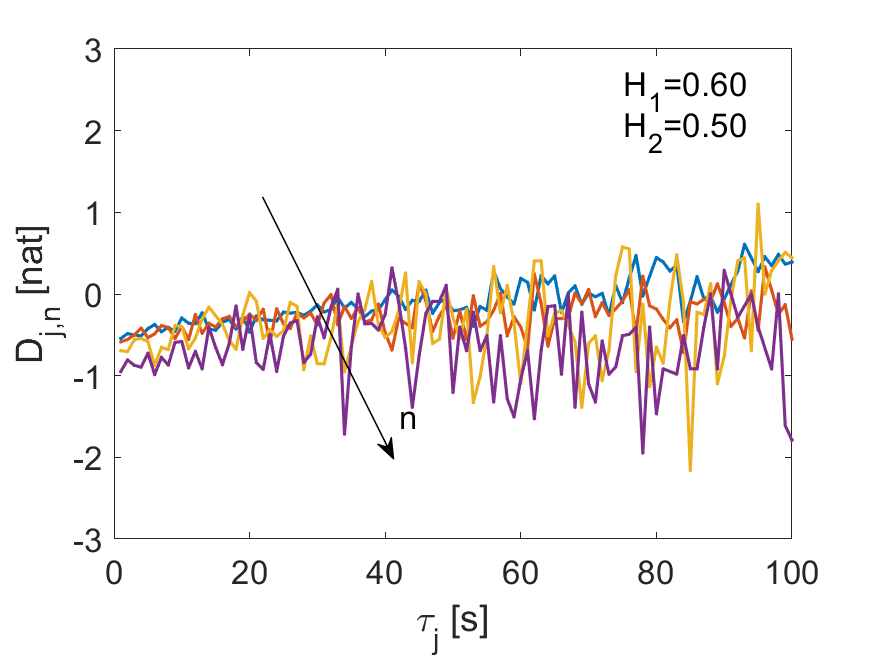}
\includegraphics[scale=0.3]{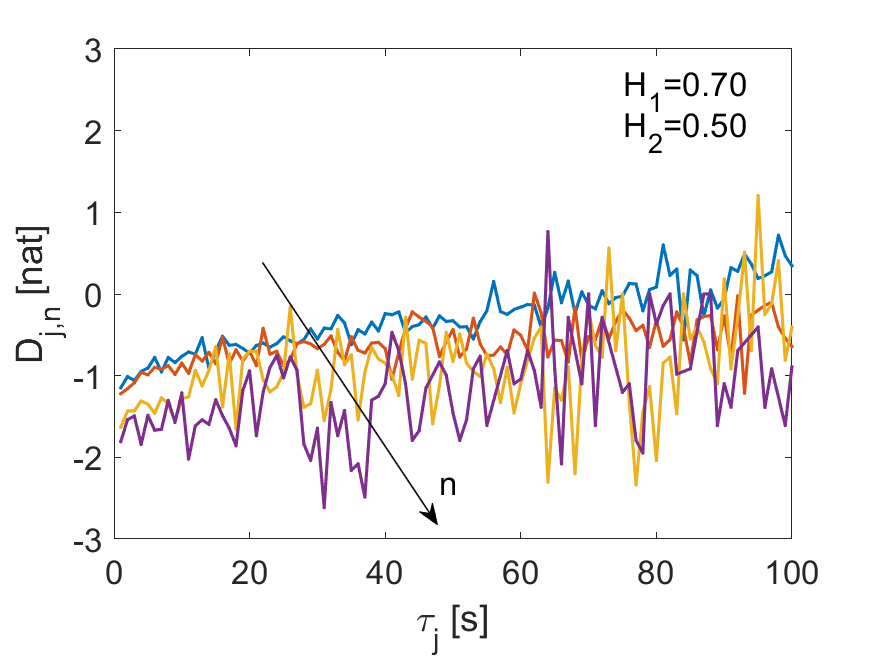}
\includegraphics[scale=0.3]{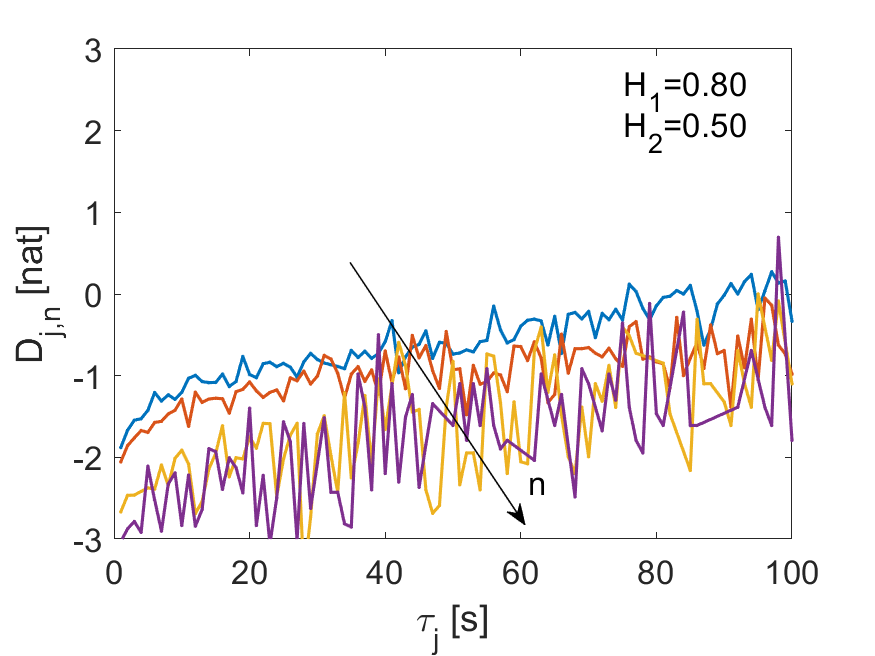}
\caption{Plot of the quantity $\mathcal{D}_{j,n}$, defined by Eq.~(\ref{Kullbackdtau}),  as a function of  the cluster duration $\tau_j$ for pairs of \textit{fBm}  with Hurst exponent $H_{1}$ and $H_{2}$. The cluster frequency $P(\tau_j,n)$ is obtained  by counting the  occurrences of the clusters with duration $\tau_j$ in  fractional Brownian motions with Hurst exponent $H_{1}$. A simple Brownian motion,  i.e. a $fBm$ with $H_{2} =0.50$, has been taken to obtain the cluster partition and  the model probability $Q(\tau_j ,n)$.  In the above figures, $H_{1}$ varies respectively from $0.20 $ (top-left)  to $0.80 $ (bottom-right). The length of the series is equal to $N=500000$ for all the graphs. Different curves in each graph refer to different $n$  values ($n=50, 100, 1000, 2000$) as indicated by the arrow. At large values of the parameter $n$, the curves tend to the asymptotic value $\mathcal{D}_{j,n} =0 $,  expected at large  $\tau_j$, whereas the curves exhibit a diverging behavior at small values of $\tau_j$. Conversely, at small values of the parameter $n$, the curves tend to the theoretical  value  expected at small values of $\tau_j$, whereas the curves diverge at large $\tau_j$. The properties of  $\mathcal{D}_{j,n}$ are discussed in Section \ref{sec:Discussion} on the basis of the analytical expression derived for continuous random variables.
  }
\label{fig:KulbackFBM}
\end{figure*}
\par
\section{Methods and Results}
\label{sec:Relative}
In this section,  the main steps of the \textit{relative cluster entropy}  approach are described.  The interest is towards the development of a divergence measure able to evaluate the situation where  a model probability distribution $Q$ is defined  in parallel to the true probability distribution function $P$ of the cluster partition.
Before illustrating  how the proposed \textit{cluster  divergence}  works, a few definitions are recalled.
\begin{figure*}[htb!]
\includegraphics[scale=0.3]{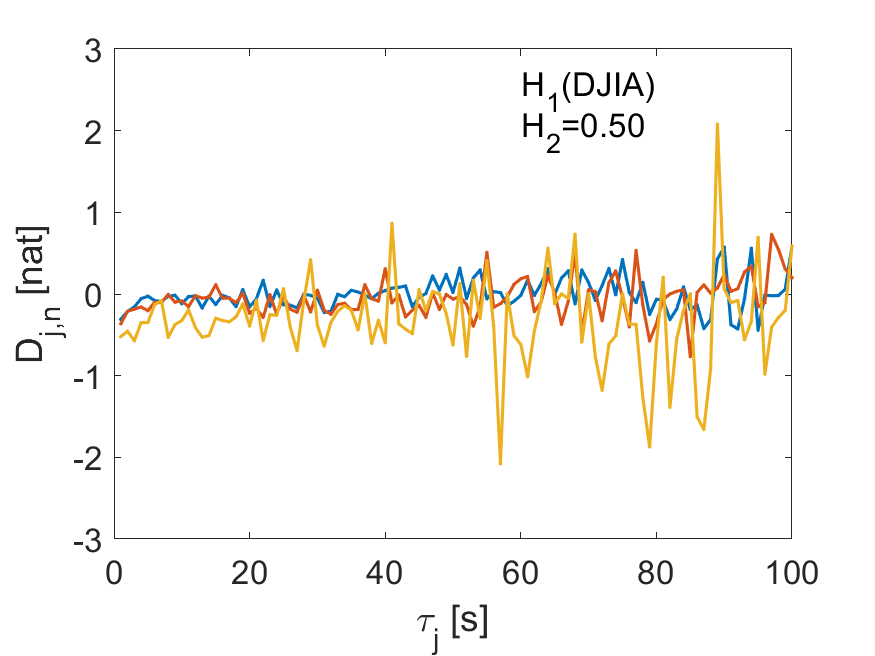}
\includegraphics[scale=0.3]{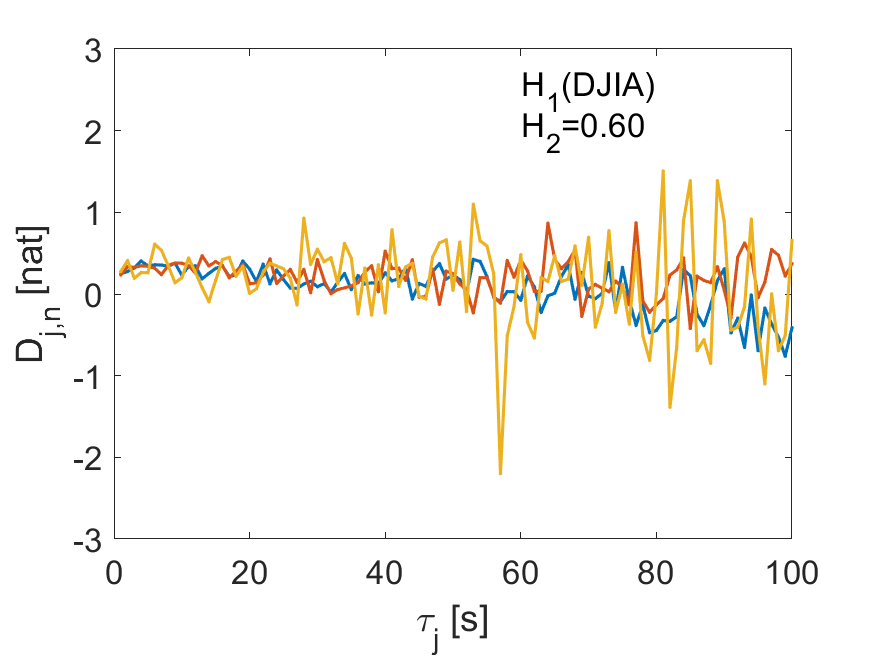}
\includegraphics[scale=0.3]{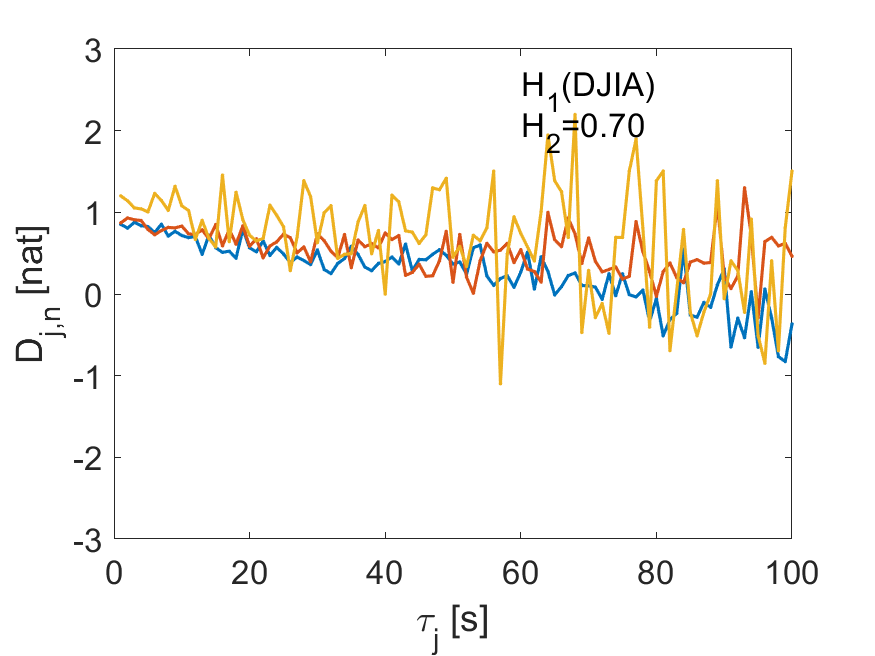}\\
\includegraphics[scale=0.3]{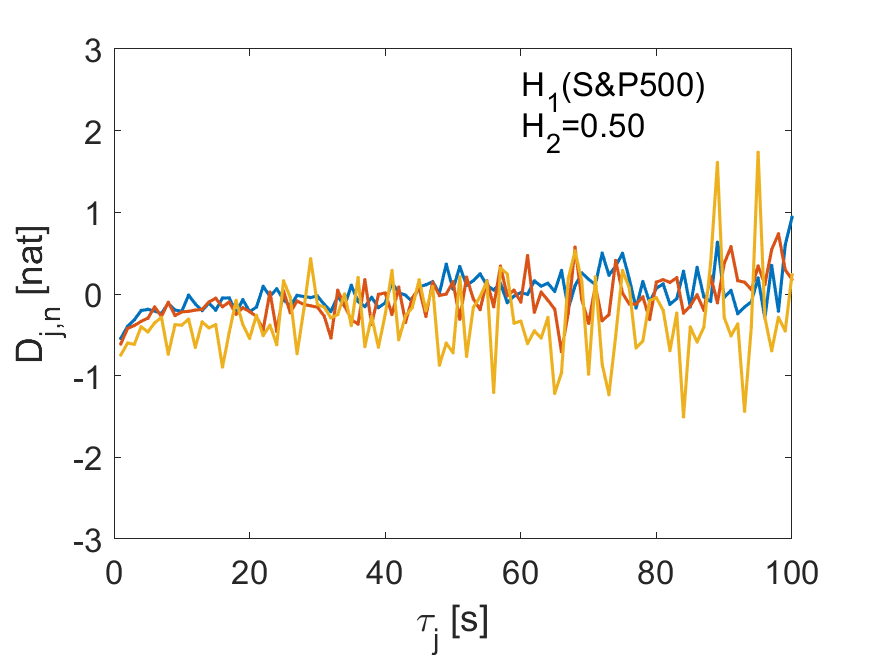}
\includegraphics[scale=0.3]{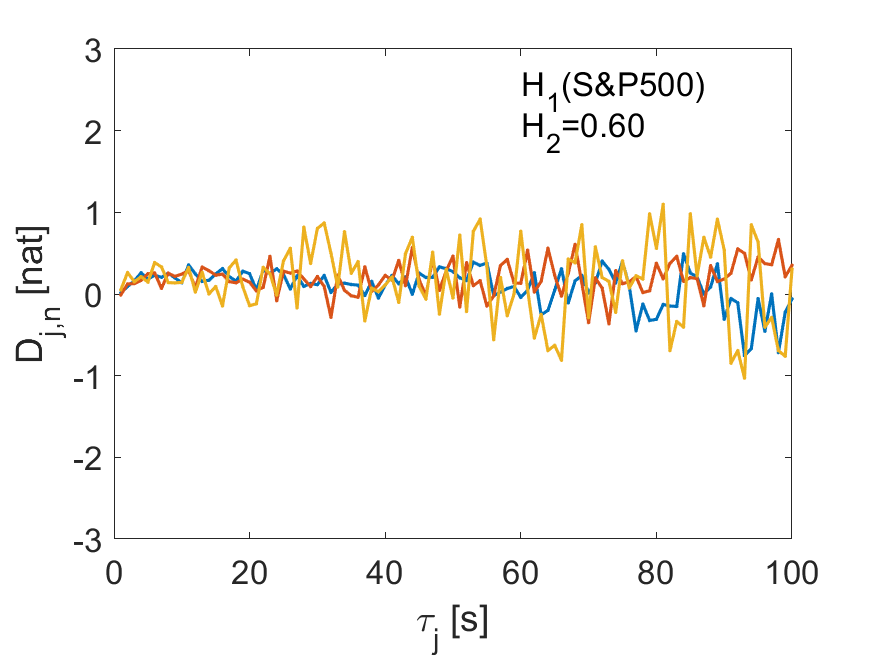}
\includegraphics[scale=0.3]{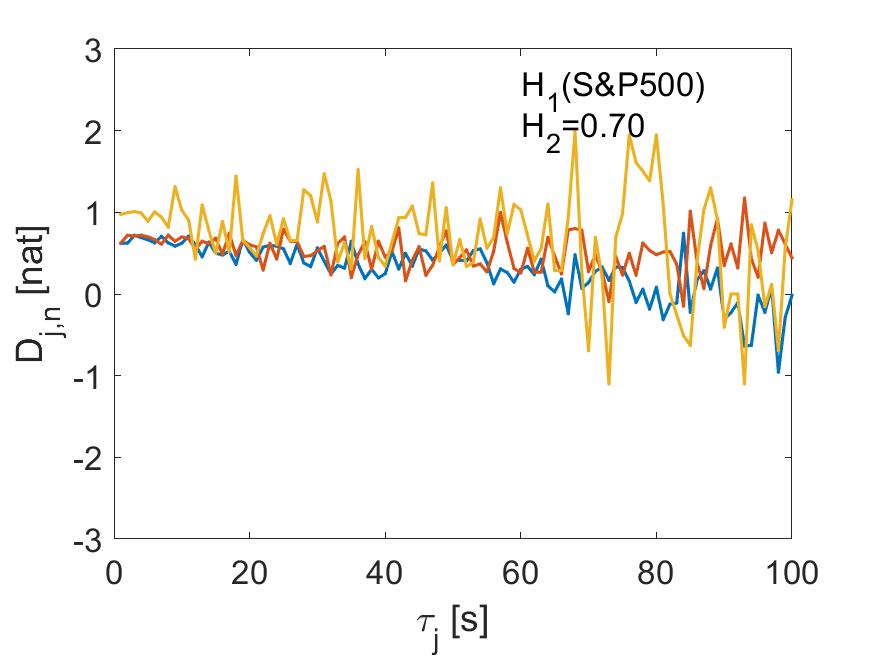}\\
\includegraphics[scale=0.3]{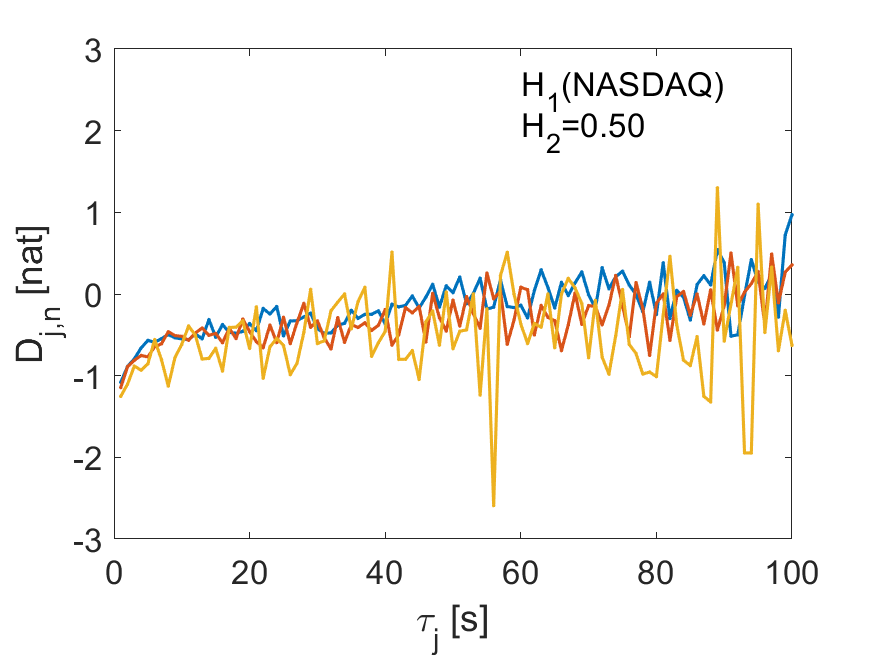}
\includegraphics[scale=0.3]{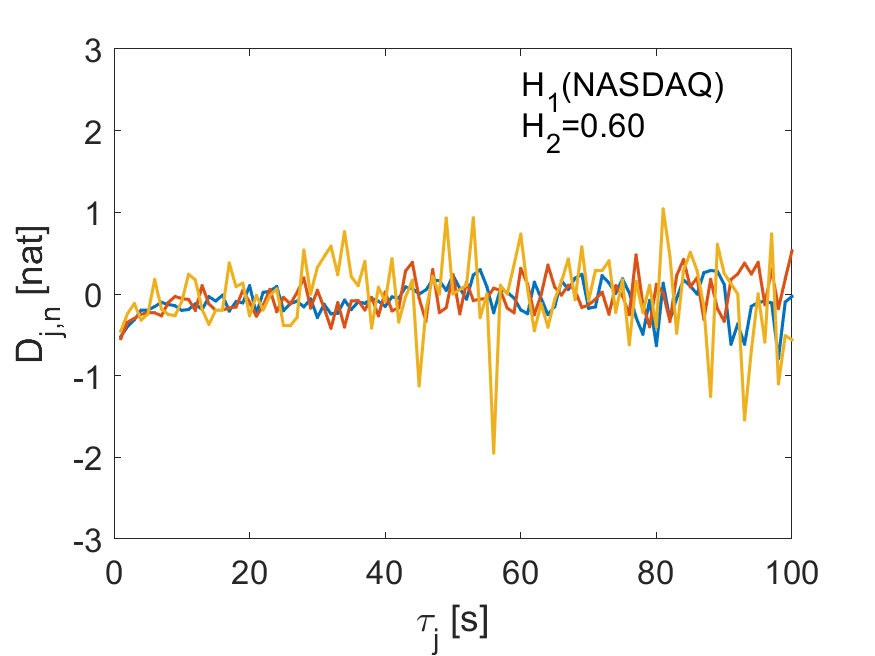}
\includegraphics[scale=0.3]{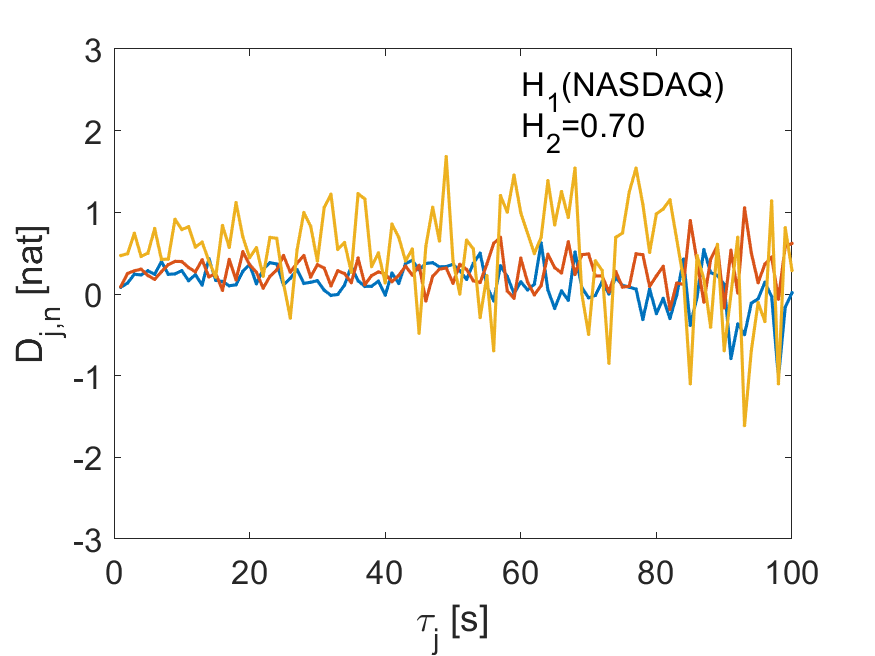}\\
\caption{ Plot of the quantity $\mathcal{D}_{j,n}$,  defined by Eq.~(\ref{Kullbackdtau}),    vs. cluster duration $\tau_j$ . The cluster frequency $P(\tau_j, n)$ has been estimated on the clusters generated by the prices of the DJIA, S\&P500,  NASDAQ indexes.   The model probability $Q(\tau_j, n)$ has been estimated on the clusters generated by a $fBm$ with Hurst exponent $H_2$ ranging from  $0.50$ to $0.70$ and length $N=492023$ equal to the length of the sampled indexes data. Different curves in each graph refer to different values of the parameter $n$ (respectively $n=50$ blue; $n=100$ orange; $n=1000$ yellow). }
\label{fig:KulbackFIN}
\end{figure*}

Consider the time series  $\{x_t \}$ of length $N$ and the local average $\widetilde{x}_{t,n} = \frac{1}{n} \sum_{n' = 0}^{n - 1} x(t-n') $   of length $N-n$  with $n \in(1,N)$.
For each $n$, a partition  $\{\cal{C}\}$  of non-overlapping clusters is generated between consecutive intersections of $\{x_t \}$ and
$\{\widetilde{x}_{t,n}\} $ defined by the time instances which make the error $\epsilon_{t,n}={x_{t}}- \widetilde{x}_{t,n}$  equal to zero. Hence, each cluster $j$ is characterized by the random variable
$\tau_j\equiv  \|t_{j}-t_{j-1}\|$,
with the instances $t_{j-1}$ and $t_j$ referring to subsequent  intersection pairs. The random variable
$\tau_j$ is named as the \textit{cluster duration}.
The empirical distribution of the    cluster duration frequencies $P(\tau_j,n)$  can be obtained by ranking the number of clusters ${\mathcal N}(\tau_1,n),{\mathcal N}(\tau_2,n), ..., {\mathcal N}(\tau_j,n)$ according to their duration $\tau_1, \tau_2,..., \tau_j$ for each $n$  as:
\begin{equation}
P(\tau_j,n)=\frac{{\mathcal N}(\tau_j,n)}{{\mathcal N_C}(n)}
\end{equation}
with  ${\mathcal N_C}(n)=\sum_{j=1}^{k(n)}  {\mathcal N}(\tau_j,n)$  the  number of clusters generated by the partition for each $n$, $k=\sum_{n=1}^{N}{\mathcal N_C}(n)$ the total number of clusters for all the possible values of $n$, and the normalization condition holding as usual:
\begin{equation}
\sum_{n=1}^N \sum_{j=1}^{{\mathcal N_C}(n)} P(\tau_j,n)= 1 \hspace{5pt}.
\end{equation}
\par
The \textit{cluster entropy} is defined as:
\begin{equation}
\mathcal{S_{C}}[P] = - \sum_{j, n} P(\tau_j, n)\log P(\tau_j,n) \hspace{5pt},
\label{Shannon}
\end{equation}
which is obtained by introducing the cluster frequency  $P(\tau_j,n)$  in the Shannon functional.
\par
In this work, the  \textit{relative cluster entropy} or  \textit{cluster divergence} $\mathcal{D_{C}}[P \| Q] $  is proposed   to quantify the wrong information yield when a model probability distribution $Q$ is assumed  in place of the empirical probability distribution $P$.
A measure of distinguishability between two probability distributions $P$ and $Q\,$ is the \textit{Kullback-Leibler divergence}, defined for discrete variables as
$ \mathcal{D_{KL}}[P\| Q]=\sum_{j} P_{j} \log \left({P_{j}}/{Q_{j}}\right) $,
with the conditions $\mathrm{supp} (P) \subseteq \mathrm{supp}(Q)$  and $ \mathcal{D_{KL}}[P\| Q] \geq 0 $, with $ \mathcal{D_{KL}}[P\| Q] = 0 $ for $P=Q\,$.
Then, the  \textit{minimum relative entropy} principle  can be adopted as  optimization criterion for model selection and statistical inference.

The  quantity $\mathcal{D}_{j,n}[P || Q ]$ is defined for each macrostate in terms of the cluster durations  $\tau_j$  as follows:
\begin{equation}
\mathcal{D}_{j,n}[P || Q ] =    P(\tau_j, n)\log \frac{P(\tau_j, n)}{Q(\tau_j, n)} \hspace{7pt},
\label{Kullbackdtau}
\end{equation}
where the index $j$  refers to the set of clusters with duration $\tau_j$ generated by the partition  for  a given $n$.
The cluster frequencies   $P(\tau_j,n)$ and $Q(\tau_j,n)$  satisfy the condition $\mathrm{supp} (P) \subseteq \mathrm{supp}(Q)$.
By using Eq.~(\ref{Kullbackdtau}) and summing $\mathcal{D}_{j,n}[P || Q ]$ over all the accessible cell states , the  \textit{relative cluster entropy} is written as:
\begin{equation}
\mathcal{D_{C}}[P|| Q]   = \sum_{n=1}^N \sum_{j=1}^{{\mathcal N_C}(n)} P(\tau_j, n)\log \frac{P(\tau_j, n)}{Q(\tau_j, n)} \hspace{5pt},
\label{Kullbackdtaun}
\end{equation}
where the index $j$ runs over the clusters obtained by each partition with size $n$, which in turn  runs over the allowed set of time window values, $n \in(1,N)$.
\begin{figure*}[]
\includegraphics[scale=0.31]{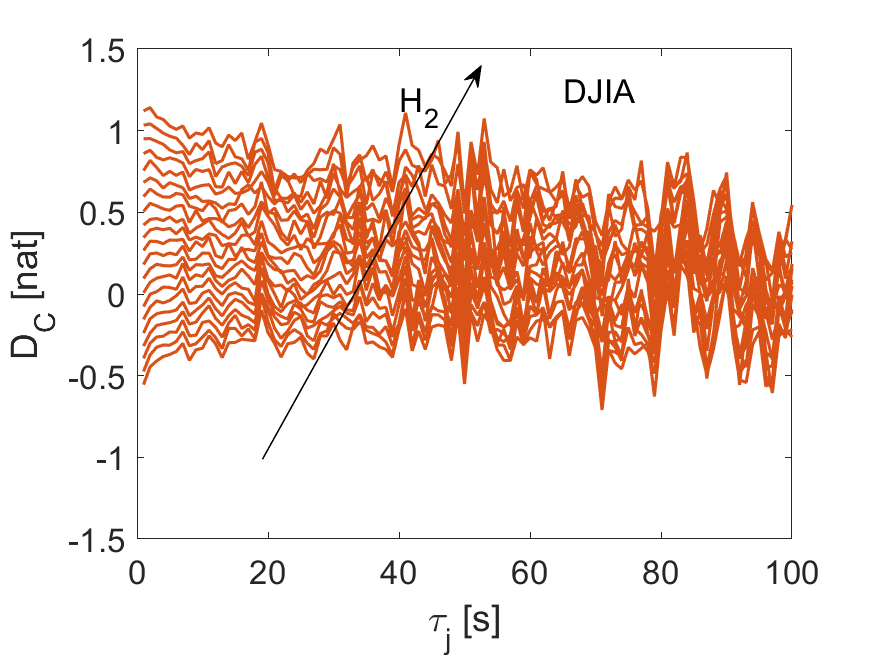}
\includegraphics[scale=0.31]{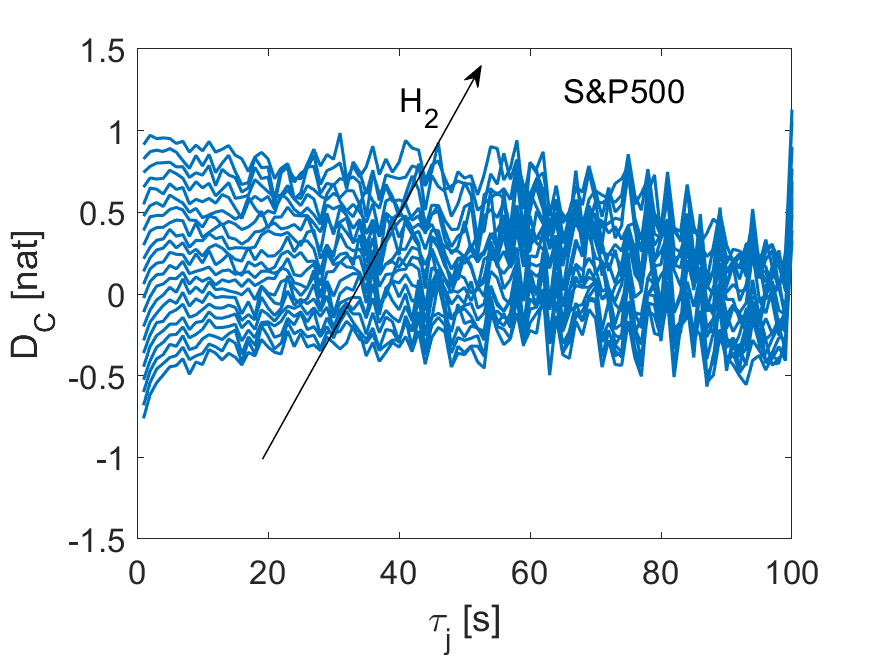}
\includegraphics[scale=0.31]{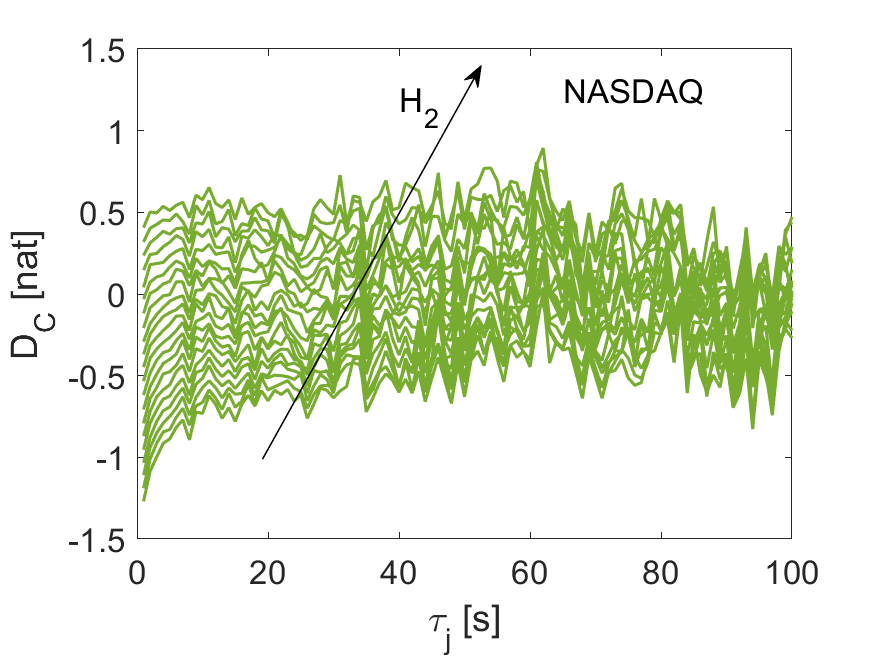}
\caption{Plot of the quantity $\mathcal{D_{C}}[P|| Q]$,  defined by Eq.~(\ref{Kullbackdtaun}), vs. cluster duration $\tau_j$ . The curves are obtained by summing the quantities $\mathcal{D}_{j,n}[P|| Q]$, as those shown in Fig.~\ref{fig:KulbackFIN}, over the parameter $n$ for the prices of DJIA, S\&P500,  NASDAQ. Each curve in the figures corresponds to the cluster divergence with the probability $P(\tau_j, n)$ referred to the market price series  $p_t$ and the model probability $Q(\tau_j, n)$ referred to \textit{fBms} with Hurst exponent $H_2$ ranging from $0.50$ to $0.70$ with step $0.1$ as indicated by the arrow. }
\label{fig:Kulback0507}
\end{figure*}
\begin{figure*}[]
\includegraphics[scale=0.4]{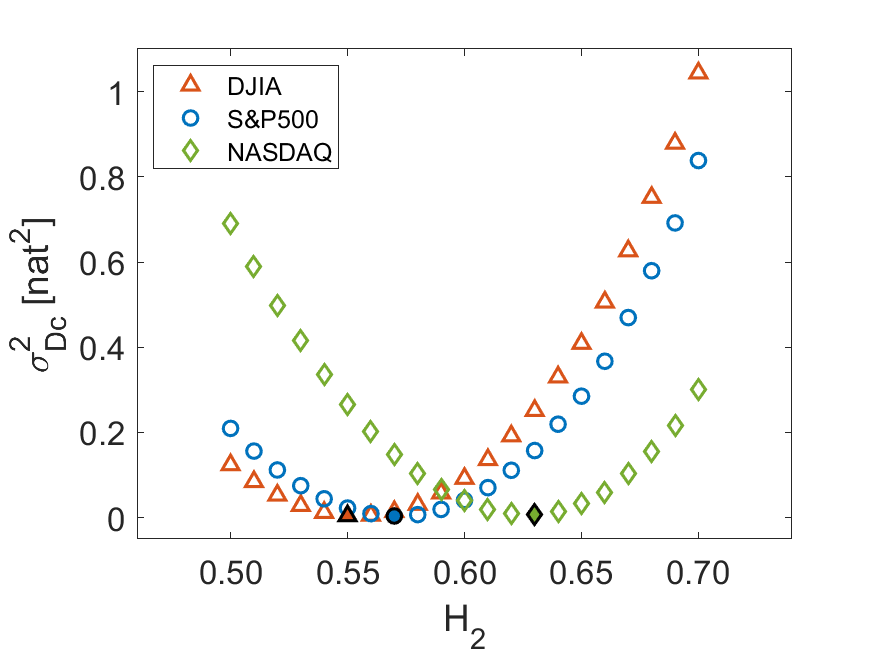}
\caption{Plot of  the quantity ${\sigma^2_{\mathcal{D}_C}} $ defined by Eq.~(\ref{Kullbackdmin}) for the relative cluster entropy  curves plotted in Fig.~\ref{fig:Kulback0507} vs. the Hurst exponent $H_2$ of the model distribution $Q(\tau_j,n)$. The quantity ${\sigma^2_{\mathcal{D}_C}} $ is the variance of $\mathcal{D}_{C}$ with respect to  $0$ (the null hypothesis for $P=Q$) over the cluster lifetime interval $1<\tau_j<20$.
Each point is evaluated by using the definition given in Eq.~(\ref{Kullbackdmin}) for each market and for each $fBm$ with assigned Hurst exponent $H_2$.
The Hurst exponent $H_2$ of the model distribution $Q(\tau_j,n)$ ranges between $0.50$ and $0.70$ with step $0.01$.  The minimum value of the variance is obtained for $H_1=0.55$ (DJIA), $H_1=0.57$  (S\&P500) and $H_1=0.63$  (NASDAQ).}
 \label{fig:varianza}
\end{figure*}
\par
To exemplify how the relative cluster entropy could  be applied in practice,  pairs of  artificially generated fractional Brownian motions (\textit{fBms}) are analysed in terms of the \textit{relative  cluster entropy} defined by Eqs.~(\ref{Kullbackdtau}-\ref{Kullbackdtaun}).  Fractional Brownian motions (\textit{fBms})  $x_t^{H} $  with ${t \geqslant 0}$ are power-law correlated stochastic processes,  defined by a centered Gaussian process with stationary increments and covariance given by $ {<x_{s}^{H}x_{t}^{H}>}=\frac{1}{2}\left(t^{2 H}+s^{2 H}-|t-s|^{2 H}\right)$ with $H \in(0,1)$ the Hurst exponent. Power-law behaviour of  the correlation function implies  slow memory decay and non-Markovianity. Synthetic \textit{fBm} sequences  have been generated with assigned Hurst exponent $H$  and  length $N$ by using the FRACLAB code \cite {fraclab}. The cluster frequencies $P(\tau_j, n)$ and $Q(\tau_j, n)$  have been estimated by counting the number of clusters with duration $\tau_j$ and window $n$ for each $fBm$.
\par
Fig.~\ref{fig:KulbackFBM} shows a few examples of plots of the quantity $\mathcal{D}_{j,n}$, defined by Eq.~(\ref{Kullbackdtau}). $\mathcal{D}_{j,n}$ is estimated  for  cluster frequency $P$, obtained from clusters generated in $fBms$ with $H_{1}$ varying from  $0.20$ (top-left)  to  $0.80 $ (bottom-right) with step $0.05$, and  model distribution $Q$ obtained from uncorrelated Brownian paths, i.e. $fBms$ with  $H_{2} =0.50$.  The values of the Hurst exponents  correspond respectively to   correlation exponents  $\alpha_1=2-H_1$ ranging from  $1.80$ to $1.20$, whereas  $\alpha_2=2-H_2$ is kept constant and equal to $1.50$.
The quantity $\mathcal{D}_{j,n}$ shows characteristic deviations with respect to the null hypothesis  corresponding to a fully random process with $H_2=0.5$. In particular,  at small values of the cluster duration $\tau_j$,  the quantity $\mathcal{D}_{j,n}$  takes  positive and negative values respectively for \textit{fBms} with $0.5<H_1<1$ and $0<H_1<0.5$. As the cluster duration $\tau_j$ increases,  $\mathcal{D}_{j,n}$   tends to reach  the horizontal axis implying that the divergence between the distributions become negligible for very large clusters.
Note in particular the three panels of the middle row in Fig.~\ref{fig:KulbackFBM}  showing the results obtained  for  fractional Brownian motions with $H_{1}=0.45 $,  $H_{1} =0.50$  and $H_{1}=0.55 $  with respect to the simple Brownian path, i.e.  the $fBm$ with  $H_{2} =0.50$, taken as the model. Thus, \textit{fBm} pairs with close values of $H_{1}$ and  $H_{2}$  correspond to more realistic experimental conditions.  Inference problems with  data sequences featuring correlation exponents  statistically close to each other  and small deviations from the model distribution should be reasonably expected in the cases of practical interest.
\par
To further illustrate how the proposed method operates with real-world data,  price series $\{p_t\} $ of  Dow Jones Industrial Average (DJIA), Standard and Poor 500 (S\&P500),  National Association of Securities Dealers Automated Quotations Composite (NASDAQ),  are considered.
Data include tick-by-tick prices  from January to  December 2018. Details (Ticker; Extended name; Country; Currency; Members; Length) provided by Bloomberg  \cite{bloomberg}. Raw data prices $\{p_t\} $ have different lengths ($N_{DIJA} = 5749145$,  $N_{S\&P500} = 6142443$,  $N_{NASDAQ} = 6982017$). To  perform the relative cluster entropy analysis over comparable data sets, raw data prices are sampled to yield equally spaced data sequences with equal length $N$. The cluster frequency $P(\tau_j,n)$  is estimated by counting the clusters generated in the market price series.  $Q(\tau_j,n)$ is estimated by counting the clusters generated in synthetic stochastic processes assumed as a model. In this analysis,  the divergence between each price series, with unknown correlation exponent, and artificially generated samples of  fractional Brownian motions \textit{fBms} with assigned Hurst exponent $H_2$, is considered. Results of the analysis  are plotted in Fig.~\ref{fig:KulbackFIN},  showing the  relative cluster entropy   for  the three markets.
Several samples of the divergence obtained for different values of the parameter $n$, shown in Fig.~\ref{fig:KulbackFIN},  have been summed over the parameter $n$, with same interval of cluster duration $\tau_j$.  Fig.~\ref{fig:Kulback0507}  shows the relative cluster entropy  $\mathcal{D_{C}}[P|| Q]$ for the data shown in Fig.~\ref{fig:KulbackFIN}.
\par
To infer the optimal probability distribution $P$, the \textit{minimum relative entropy} principle is implemented non-parametrically on the  values plotted in Fig.~\ref{fig:Kulback0507}. To this purpose, the variance $\sigma^2_{\mathcal{D}_C} $ of  $\mathcal{D}_{C}[P|| Q ] $ around the value $\mathcal{D}_{C}[Q||Q] $ (the null hypothesis  for $P=Q$) is written as follows:
\begin{equation}
{\sigma^2_{\mathcal{D}_C}} \equiv \frac{1}{k-1} \sum_{j=1}^k
\left[\mathcal{D}_{C}[P|| Q]- \mathcal{D}_{C}[Q||Q]\right]^2
\label{Kullbackdmin}
\end{equation}
where the sum  runs over the total number of clusters obtained by the partition process. By using the value $\mathcal{D}_{C}[Q||Q] =0$, Eq.~(\ref{Kullbackdmin}) writes:
\begin{equation}
{\sigma^2_{\mathcal{D}_C}} = \frac{1}{k-1}  \sum_{j=1}^k
\left[\mathcal{D}_{C}[P|| Q]\right]^2
\end{equation}
The quantity ${\sigma^2_{\mathcal{D}_C}}$ corresponds to the mean square value of the area of the region between the curve $\mathcal{D}_{C}[P||Q]$ and the horizontal axis ($\mathcal{D}_{C}[Q||Q]=0$). Given the linearity of the relative cluster entropy operator,  $\sigma^2_{\mathcal{D}_C} $ exhibits a quadratic behaviour with the typical  asymmetry of the Kullback-Leibler entropy. The quadratic functional can be easily used to estimate the minimum.
\begin{figure*}[htb!]
\includegraphics[scale=0.4]{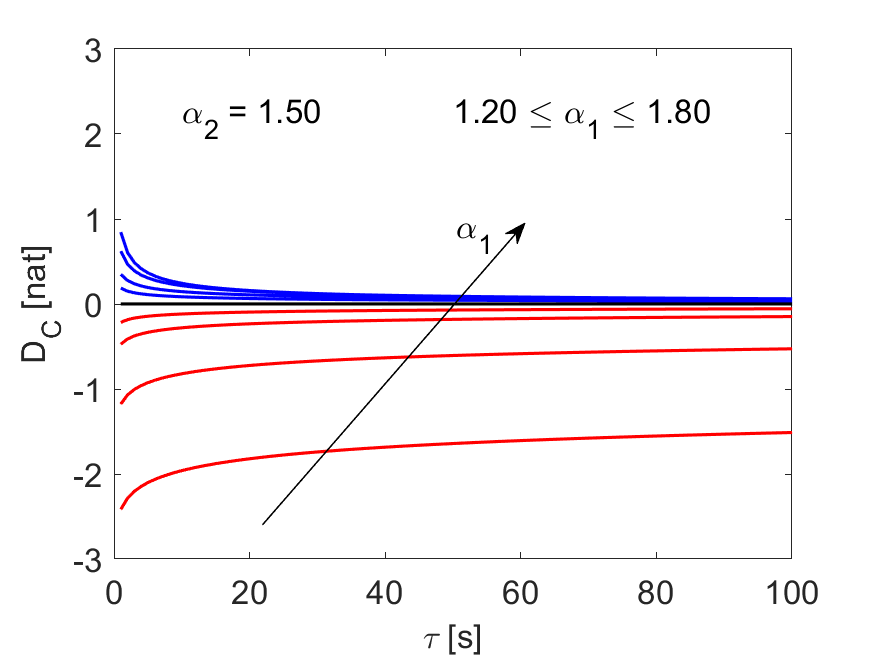}
\caption{Plot of  the quantity $\mathcal{D_{C}}[P|| Q]$, defined by  Eq.~(\ref{Kullbackc0}), vs. the cluster duration $\tau$.  Blue curves correspond to a power law probability distribution $P(\tau)$ with ${\alpha_1}$  ranging between $1.55 \div 1.80$.  The model probability distribution $Q(\tau)$ is a power law with correlation  exponent ${\alpha_2} =1.50 $, the same for all the curves plotted here.  Red curves correspond to a power law probability distribution with  ${\alpha_1}$  ranging between $1.20 \div 1.45$. The black line corresponds to the null hypothesis $\mathcal{D_{C}}[P|| P] =0$ obtained with  ${\alpha_1}=1.50$ and ${\alpha_2} =1.50$.
}
\label{fig:Kulbackth}
\end{figure*}
The minimization criterion provided by   Eq.~(\ref{Kullbackdmin}) has been applied to the data shown in  Fig.~\ref{fig:Kulback0507} to yield the best estimate of the correlation degree of the market prices.  The value of the Hurst exponent for the series of the prices $\{p_t\}$ has been deduced from the value of $H_2$  for which $\sigma^2_{\mathcal{D}_C} $ takes its minimum,  implying $H_1=H_2$.  By using this rule, $H_1=H_2=0.55$,  $H_1=H_2=0.57$,   and $H_1=H_2=0.63$  have been found respectively  for  DJIA, S\&P500  and NASDAQ.
The minimization outcomes are plotted in Fig.~\ref{fig:varianza}  for the markets  shown in  Fig.~\ref{fig:Kulback0507}.
\section{Discussion and Conclusion}
\label{sec:Discussion}
In this Section,   the  \textit{relative cluster entropy} is extended to continuous  random variables.
For $N_{\mathcal{C}} (n) \rightarrow \infty$, the characteristic size of generated clusters $\cal{C}$   behaves as continuous random  variables $\tau \in [1, \infty]$ with  probability distribution function $P(\tau)$ varying as a power-law \cite{carbone2004analysis,carbone2007scaling}.
 By taking the limits
$P({\tau_j}) \rightarrow P(\tau) d \tau$
and $Q ({\tau_j}) \rightarrow Q(\tau) d \tau$,   Eq. (\ref{Kullbackdtaun})  can be written for continuous random variables in the form of an integral:
\begin{equation}
{D_{C}}[P(\tau)|| Q(\tau)]= \int P(\tau)\log \frac{ P\left({\tau}\right)}{Q\left({\tau}\right)} d \tau \hspace{5pt},
\label{Kullbackctau}
\end{equation}
with $\tau \in [1, \infty]$.
We are interested in the situations where the probability distributions    are  power-law functions, i.e. for $P(\tau)$ and $Q(\tau)$ respectively in the form:
\begin{equation}
 P(\tau)=(\alpha_1-1) \tau^{-\alpha_1}  \hspace{20pt}  Q(\tau)= (\alpha_2 -1)\tau^{-\alpha_2}   \hspace{5pt},
\label{PQ}
\end{equation}
where ${\alpha_1}$ and  ${\alpha_2}$ are the correlation exponents, $ \alpha_1-1$  and  $ \alpha_2 -1$ are the normalization constants for  $\tau \in [1, \infty]$.
By using Eqs.~(\ref{PQ}),  Eq.~(\ref{Kullbackctau}) writes:
\begin{equation}
{D_{C}}[P(\tau)|| Q(\tau)] = \int  (\alpha_1-1)  \tau^{-\alpha_1} \log \frac{ (\alpha_1-1)  \tau^{-\alpha_1}}{(\alpha_2 -1)\tau^{-\alpha_2} } d \tau  \hspace{5pt},
\end{equation}
that after integration becomes:
\begin{widetext}
\begin{equation}
{D_{C}}[P(\tau)|| Q(\tau)] =  \tau^{1-\alpha_1} \left (\log \frac {\alpha_1-1}{\alpha_2-1}
+   \left (\log {\tau^{(\alpha_1 -\alpha_2)}} +\frac{\alpha_1 -\alpha_2 }{1-\alpha_1}\right) \right)+ \hspace{2pt} \text{constant}  \hspace{3pt},
\label{Kullbackc0}
\end{equation}
\end{widetext}
where the integration constant  is equal to zero by setting  ${D_{C}}[P|| P]=0$.
By estimating  the definite integral  over the interval  $[1, \infty]$, one obtains:
\begin{equation}
{D_{C}}[P|| Q] =    \log \frac {\alpha_1-1}{\alpha_2-1}
+   \frac{ \alpha_1 -\alpha_2 }{1- \alpha_1} \hspace{5pt},
\label{Kullbackc00}
\end{equation}
  that for  $\alpha_1 = \alpha_2\,$, i.e. for the distribution $P$ coincident with the model distribution $Q$, provides ${D_{C}}[P||Q] = 0$.
\par
${D_{C}}[P|| Q]$ quantifies the divergence between $P(\tau)$ and $Q(\tau)$, respectively true and model distribution,  as a function of the cluster lifetime $\tau$  in terms of the pair of correlation exponents   $\alpha_1$  and $\alpha_2$.   Eq.~(\ref{Kullbackc0}) is plotted as a function of $\tau$  for  different values of the exponents $\alpha_1$ and $\alpha_2$ in Fig.~\ref{fig:Kulbackth}.
At small values of the cluster duration ($\tau \rightarrow 1$),  ${D_{C}}[P|| Q]$  is strongly dependent on the difference of the power-law exponent $\alpha_1$ with respect to the exponent $\alpha_2$ of the model distribution. Conversely, as   the cluster duration increases ($\tau >> 1$),   ${D_{C}}[P|| Q]$  becomes negligible. The decay  can be understood by considering that as  $\tau$ increases the cluster becomes disordered as a consequence of the spread of the distribution and the onset of finite-size effect. The  correlation vanishes as the process becomes almost fully uncorrelated.
The behaviour  of the cluster distribution divergence obtained by using continuous variables is  consistent with the empirical tests performed on discrete data sets. In particular, the behaviour shown  by  the fractional Brownian motions with different correlation exponents discussed in the  Section II is reproduced by the curves shown in  Fig.~\ref{fig:Kulbackth}, ensuring that the approach is sound and robust.
\par
The  \textit{relative cluster entropy}  can be therefore exploited to estimate the deviation of the power law exponent corresponding respectively to experimental and model probability distributions.
\par
Long-range correlated processes obeying power-law distributions occur frequently in complex system data related to several  natural and man-made phenomena.  Due to their ubiquity, the  extent of long-range correlation and the scaling exponents are relevant to many disciplines, though several difficulties are met for their estimation  which require suitable computational procedures to be carefully implemented \cite{clauset2009power}.
A random variable $x$ obeys a power law if it is drawn from a probability distribution
$p(x) \propto x^{-\alpha}$
with  $\alpha>1$ the correlation exponent.
Empirical real-world data barely follow a power-law for all the values of $x$. Due  to normalization requirements and finite-size effects, ideal power-law behaviour usually holds at values greater than some minimum $x_{\min}$ up to a maximum $x_{\max}$. An exponential cut-off is often artificially introduced to account for the deviation from the ideal power-law behaviour
$x^{-\alpha} \mathrm{e}^{-\lambda x}$.

\par
The non-parametric minimization of the relative entropy has some  advantages compared to the parametric approaches, whose implementation requires normality of the random variables and knowledge of the first two moments of the distribution for the calculation of the Lagrange multipliers.
 The proposed \textit{relative cluster entropy} approach yields the optimal value of the correlation exponent $\alpha$ without relying on the estimate of the slope in a log-log plot.  Thus the proposed approach is robust against computational biases which usually affect least-squares estimates.

\bibliographystyle{unsrt}

\begin{thebibliography}{10}

\bibitem{cafaro2016thermodynamic}
Carlo Cafaro, Sean~Alan Ali, and Adom Giffin.
\newblock Thermodynamic aspects of information transfer in complex dynamical
  systems.
\newblock {\em Physical Review E}, 93(2):022114, 2016.

\bibitem{parrondo2015thermodynamics}
Juan~MR Parrondo, Jordan~M Horowitz, and Takahiro Sagawa.
\newblock Thermodynamics of information.
\newblock {\em Nature physics}, 11(2):131--139, 2015.

\bibitem{kawai2007dissipation}
Ryoichi Kawai, Juan~MR Parrondo, and Christian Van~den Broeck.
\newblock Dissipation: The phase-space perspective.
\newblock {\em Physical review letters}, 98(8):080602, 2007.

\bibitem{horowitz2014thermodynamics}
Jordan~M Horowitz and Massimiliano Esposito.
\newblock Thermodynamics with continuous information flow.
\newblock {\em Physical Review X}, 4(3):031015, 2014.

\bibitem{still2012thermodynamics}
Susanne Still, David~A Sivak, Anthony~J Bell, and Gavin~E Crooks.
\newblock Thermodynamics of prediction.
\newblock {\em Physical review letters}, 109(12):120604, 2012.

\bibitem{ortega2013thermodynamics}
Pedro~A. Ortega and Daniel~A. Braun.
\newblock Thermodynamics as a theory of decision-making with
  information-processing costs.
\newblock {\em Proceedings of the Royal Society A: Mathematical, Physical and
  Engineering Sciences}, 469(2153):20120683, 2013.

\bibitem{san2005information}
X~San~Liang and Richard Kleeman.
\newblock Information transfer between dynamical system components.
\newblock {\em Physical review letters}, 95(24):244101, 2005.

\bibitem{chen2021wiener}
Junya Chen, Jianfeng Feng, and Wenlian Lu.
\newblock A wiener causality defined by divergence.
\newblock {\em Neural Processing Letters}, 53(3):1773--1794, 2021.

\bibitem{vedral2002role}
Vlatko Vedral.
\newblock The role of relative entropy in quantum information theory.
\newblock {\em Reviews of Modern Physics}, 74(1):197, 2002.

\bibitem{kleeman2002measuring}
Richard Kleeman.
\newblock Measuring dynamical prediction utility using relative entropy.
\newblock {\em Journal of the atmospheric sciences}, 59(13):2057--2072, 2002.

\bibitem{granero2018kullback}
Carlos Granero-Belinch{\'o}n, St{\'e}phane~G Roux, and Nicolas~B Garnier.
\newblock Kullback-leibler divergence measure of intermittency: Application to
  turbulence.
\newblock {\em Physical Review E}, 97(1):013107, 2018.

\bibitem{backus2014sources}
David Backus, Mikhail Chernov, and Stanley Zin.
\newblock Sources of entropy in representative agent models.
\newblock {\em The Journal of Finance}, 69(1):51--99, 2014.

\bibitem{tozzi2021information}
Arturo Tozzi and James~F Peters.
\newblock Information-devoid routes for scale-free neurodynamics.
\newblock {\em Synthese}, 199(1):2491--2504, 2021.

\bibitem{ullmann2021validation}
Theresa Ullmann, Christian Hennig, and Anne-Laure Boulesteix.
\newblock Validation of cluster analysis results on validation data: A
  systematic framework.
\newblock {\em Wiley Interdisciplinary Reviews: Data Mining and Knowledge
  Discovery}, page e1444, 2021.

\bibitem{meilua2007comparing}
Marina Meil{\u{a}}.
\newblock Comparing clusterings—an information based distance.
\newblock {\em Journal of multivariate analysis}, 98(5):873--895, 2007.

\bibitem{liao2005clustering}
T~Warren Liao.
\newblock Clustering of time series data—a survey.
\newblock {\em Pattern recognition}, 38(11):1857--1874, 2005.

\bibitem{carbone2004analysis}
Anna~Carbone, Giuliano~Castelli, and H.~Eugene Stanley.
\newblock Analysis of clusters formed by the moving average of a long-range
  correlated time series.
\newblock {\em Physical Review E}, 69:026105, Feb 2004.

\bibitem{carbone2007scaling}
Anna~Carbone and H.~Eugene Stanley.
\newblock Scaling properties and entropy of long-range correlated time series.
\newblock {\em Physica A: Statistical Mechanics and its Applications},
  384(1):21--24, 2007.

\bibitem{carbone2013information}
Anna~Carbone.
\newblock Information measure for long-range correlated sequences: the case of
  the 24 human chromosomes.
\newblock {\em Scientific Reports}, 3:2721, 2013.

\bibitem{ponta2021information}
Linda Ponta, Pietro Murialdo, and Anna Carbone.
\newblock Information measure for long-range correlated time series:
  Quantifying horizon dependence in financial markets.
\newblock {\em Physica A: Statistical Mechanics and its Applications}, page
  125777, 2021.



\bibitem{clauset2009power}
Aaron Clauset, Cosma~Rohilla Shalizi, and Mark~EJ Newman.
\newblock Power-law distributions in empirical data.
\newblock {\em SIAM review}, 51(4):661--703, 2009.


\bibitem{fraclab}
\url{https://project.inria.fr/fraclab/}

\bibitem{bloomberg}
\url{www.bloomberg.com/professional}
\end{thebibliography}

\end{document}